\documentclass[aps,prl,twocolumn,superscriptaddress]{revtex4-1}

\usepackage{graphicx}
\usepackage{dcolumn}
\usepackage{bm}

\begin{document}


\title{Muonium emission into vacuum from mesoporous thin films at cryogenic temperatures}

\author{A.~Antognini}
\email[]{aldo@phys.ethz.ch}
\affiliation{Institute for Particle Physics, ETH Zurich, Switzerland}

%
\author{P.~Crivelli}
\email[]{crivelli@phys.ethz.ch}
\affiliation{Institute for Particle Physics, ETH Zurich, Switzerland}
%
\author{T.~Prokscha}
\email[]{thomas.prokscha@psi.ch}
\affiliation{Paul Scherrer Institute, Villigen, Switzerland}
\author{K.~S.~Khaw}
\affiliation{Institute for Particle Physics, ETH Zurich, Switzerland}%
\author{B. Barbiellini}
\affiliation{Department of Physics, Northeastern University, Boston, Massachusetts 02115, USA}
\author{L.~Liszkay}
\affiliation{CEA, Irfu, S\'edi, Centre de Saclay, F-91191 Gif-sur-Yvette, France}
\author{K.~Kirch} 
\affiliation{Institute for Particle Physics, ETH Zurich, Switzerland}
\affiliation{Paul Scherrer Institute, Villigen, Switzerland}
\author{K.~Kwuida} 
\affiliation{Institute for Particle Physics, ETH Zurich, Switzerland}
\author{E.~Morenzoni}
\affiliation{Paul Scherrer Institute, Villigen, Switzerland}
\author{F.~M.~Piegsa}
\affiliation{Institute for Particle Physics, ETH Zurich, Switzerland}
\author{Z.~Salman}
\affiliation{Paul Scherrer Institute, Villigen, Switzerland}
\author{A.~Suter}
\affiliation{Paul Scherrer Institute, Villigen, Switzerland}
\date{\today}

\begin{abstract}
  We report on Muonium (Mu) emission into vacuum following $\mu^+$
  implantation in  mesoporous thin
  SiO$_2$ films. We obtain a yield of Mu into vacuum of (38$\pm$4)\% at
  250~K temperature and (20$\pm$4)\% at 100~K for 5~keV $\mu^+$
  implantation energy.  From the implantation energy dependence of the
  Mu vacuum yield we determine the Mu diffusion constants in these
  films: $D\mathrm{^{250K}_{Mu}}=(1.6\pm 0.1)\times 10^{-4}$ cm$^2$/s and
  $D\mathrm{^{100K}_{Mu}}=(4.2\pm0.5)\times 10^{-5}$ cm$^2$/s. 
  Describing the diffusion process as quantum mechanical tunneling
  from pore-to-pore, we reproduce the measured temperature 
  dependence $\sim T^{3/2}$ of the diffusion constant. We
  extract a potential barrier of $(-0.3\pm0.1)$~eV
  which is consistent with our computed Mu work-function 
  in SiO$_2$ of $[-0.3,-0.9]$~eV. The high Mu vacuum yield even at low temperatures represents an
  important step towards next generation Mu spectroscopy experiments.
\end{abstract}

\pacs{Valid PACS appear here}
\maketitle

Muonium (Mu), the bound state of a positive muon ($\mu^+$) and an
electron, is a pure leptonic atom.
It is thus an ideal object for testing bound-state quantum
electrodynamics (QED)  free from hadronic
uncertainties related to the structure of the nucleus \cite{Jungmann,SavelyPhysRep}.
A renewed interest in this simple system has been triggered by the
recent results of the muonic hydrogen experiment~\cite{muonicH}. 
The puzzling proton radius discrepancy observed there could be ascribed to problems
related either to bound-state QED theory, or the Rydberg
constant, or the structure of the proton or new physics.
Spectroscopy of Mu addresses the first two items.
Furthermore, Mu spectroscopy provides precise determination of other
fundamental constants like the muon mass and the fine structure
constant~\cite{Muonium1S2SExp1,Muonium1S2SExp2,MuoniumHFSExp}.
Mu can also be used to search for new physics such as lepton flavor
violation via muonium-antimuonium oscillation \cite{muantimu}.

For next generation experiments, it is essential to have a source of
Mu with high vacuum yield down to low temperature and long term
stability.
Mu in vacuum is typically produced by stopping a low momentum $\mu^+$ beam
close to the surface of tungsten
foils~\cite{MuW} or silica powders~\cite{MuSiO2powder}.
The fraction of Mu which diffuses to the surface is emitted into vacuum.
Prior to this study, the highest measured vacuum yield was  $(18\pm2)\%$ per stopped $\mu^+$ obtained in SiO$_2$ powders at 300~K~\cite{MuSiO2powder,MuTracking,JanissenPhysRevA}.
Moreover, to our knowledge, Mu emission into vacuum below room
temperature has never been reported.
A Mu source with a larger flux can be achieved either by improving the $\mu^+$ beam (smaller phase space, low energy, high intensity) as proposed in~\cite{mufactories,taqqu} or by improving the $\mu^+
\rightarrow$ Mu conversion.
In this work, we focus on the optimization of the latter using
SiO$_{2}$ porous films (F-samples of \cite{oPsTOF}) which we preselected with the ETH Zurich slow positron beam. The choice of this
material was motivated by the fact that Positronium (Ps, the
electron-positron bound state) and Mu share similar formation
mechanisms. Recently, a yield of Ps into vacuum as high as 40\% from these porous samples has been measured down to cryogenic temperatures~\cite{oPsTOF}.

For this study, we used the low energy positive muon beam (LEM)
at PSI delivering approximately $3000$ s$^{-1}$ $\mu^+$ on target with energies tunable from 1 to 30~keV \cite{LEM2000,LEMNIMA}.
The $\mu^+$ are implanted in the porous film of 1~$\mu$m thickness,
pore size of (5$\pm$0.5) nm and density of 1.1 g/cm$^3$.
The mean implantation depth is 75(270)~nm for a $\mu^+$ implantation energy of 5(19)~keV.
The Mu formation mechanism is similar to the one in SiO$_2$ powders
\cite{MuSiO2powder,MuSiO2Formation}.  The $\mu^+$ implanted at keV
energy in the SiO$_2$ film rapidly thermalize in the bulk (in tens
of ps). A fraction of them forms Mu in the bulk. Those atoms diffuse until they
are ejected in the pores with almost 100\% probability
\cite{MuSiO2powder}.
The porous films have a network of interconnected pores in which
Mu can diffuse and loose its energy via collisions with the pore
walls.
If Mu reaches the film surface before decaying, it is emitted into
vacuum.
We define the Mu vacuum yield as the probability
of Mu emission into vacuum per implanted $\mu^+$.
If Mu suffered a sufficient number of collisions, during its diffusion to
the surface, it becomes thermalized at the film temperature.
While Ps from similar films is emitted into vacuum with an energy above
room temperature \cite{oPsTOF,cassidy} due to quantum mechanical
confinement in the pores, for Mu one does not expect such a limitation
because the \mbox{de Broglie} wavelength is about $10$ times smaller, i.e., of the order of 0.4 nm. 

\begin{figure}
\includegraphics[width=0.4\textwidth]{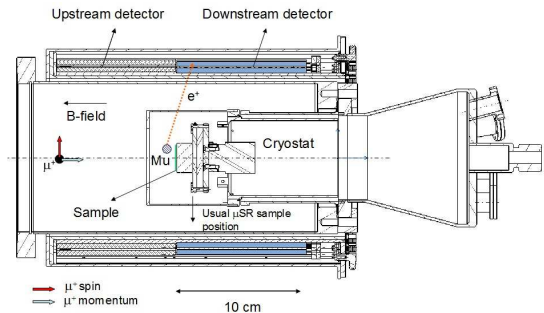}
\caption{LEM sample chamber. The sample is glued on a silver coated copper mount
  contacted to a cryostat. The sample is surrounded by scintillators for positron detection
  grouped in upstream and downstream counters. Each of them is additionally
  segmented in top, bottom, left and right.}
\label{setup}
\end{figure}

The LEM is a dedicated facility for $\mu$SR (muon spin rotation) measurements.
%
A sketch of the sample region and the positron detectors is shown in 
Fig.~\ref{setup} (see~\cite{LEM2000,LEMNIMA} for more details).
Before stopping in the sample the $\mu^+$, which are almost 100\%
transversely polarized, cross a 10~nm thin carbon foil causing the
emission of secondary electrons.
These electrons detected by a micro-channel plate provide the event
trigger.
Segmented plastic scintillators surrounding the sample region in a
cylindrical geometry are used to detect the positron from muon
decay.
The positron signal provides the stop time of the event.
The sample resides in a magnetic field transverse to the
muon spin (see Fig.~\ref{setup}).
Therefore, the muon spin undergoes Larmor precession whose frequency depends on the local magnetic field and on whether the muon remains a free $\mu^+$ or binds with an electron to form Mu.
Since the positron from muon decay is emitted preferentially along the muon spin, using a segmented detector divided in four sections (top, bottom, left, right), it is possible to track the spin precession.
The time spectra measured in each individual segment follow the
exponential  muon decay distribution, modulated at the Larmor frequency.
The number of counts $N(t)$ measured in one of the positron detectors, e.g., upstream top, is~\cite{MuSiO2powder}:
\begin{eqnarray*}
N(t)=N_0\,e^{-t/\tau}[1+A_{\mu}(t)+A_\mathrm{Mu}(t)]+B
\end{eqnarray*}
where $N_0$ is the normalization, $\tau=2.2\;\mu$s is the muon lifetime, $B$ the uncorrelated background, $A_{\mu}(t)=A_{\mu}e^{-\lambda_{\mu}t}\cos{(\omega_{\mu} t-\phi_{\mu})}$ and $A_\mathrm{Mu}(t)=A_\mathrm{Mu}e^{-\lambda_{\mathrm{Mu}}t}\cos{(\omega_{\mathrm{Mu}}t-\phi_{\mathrm{Mu}})}$ are the precession signals at frequencies  $\omega_{\mu}$ for free $\mu^+$ and
$\omega_\mathrm{Mu}$ for Mu and phases $\phi_{\mu}$ and $\phi_{\mathrm{Mu}}$. The constants $\lambda_{\mu}$ and $\lambda_{\mathrm{Mu}}$ take into account the damping of the precession signal amplitudes $A_{\mu}$ and $A_\mathrm{Mu}$ due to spin relaxation processes~\cite{MuSiO2Formation}.  
Since the gyromagnetic factor of Mu in the triplet state (F=1,
M=$\pm$1) is 103 time larger than the gyromagnetic factor of $\mu^+$ ($\omega_{\mathrm{Mu}}\approx103\,\omega_{\mu^+}$), it is
possible to clearly distinguish if an implanted $\mu^+$ remains unbound or forms Mu.
%

The initial fraction of Mu formed in the sample per implanted $\mu^+$
is determined with $F^0_\mathrm{Mu}=1-A_{\mu^+}/A_{tot}$ where
$A_{tot}$ is the total observable asymmetry which has been measured
using a Suprasil (fused quartz) reference.
The correctness of this indirect approach relies on the fact that
$\mu^+$ is not expected to depolarize in
silica~\cite{MuDepolarizationSiO2}, and thus the missing $\mu^+$
fraction is the one that converted to Mu.
\begin{figure}
\includegraphics[width=.4\textwidth]{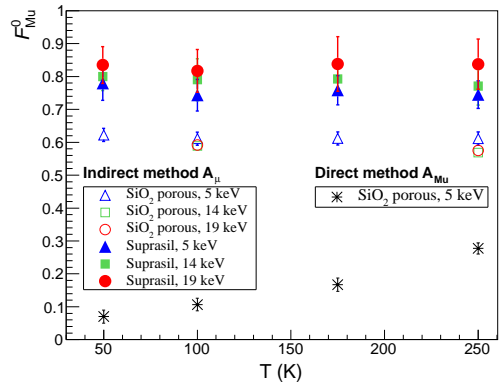}
\caption{$F^0_\mathrm{Mu}$ versus temperature for
  the porous film ($8\times 10^6$ events) and Suprasil
  ($2.5\times 10^6$ events) for various implantation
  energies obtained from the $\mu$SR amplitudes $A_\mu$ and  $A_\mathrm{Mu}$.}
\label{muSR}
\end{figure}
The measured Mu formation probability (see Fig.~\ref{muSR}) for porous
SiO$_2$ is $F^0_\mathrm{Mu}=(60\pm2)$\% which is comparable with the results obtained in
silica powders~\cite{MuSiO2powder}.
For Suprasil we obtained $F^0_\mathrm{Mu}=(80\pm4)$\% in agreement with \cite{MuSiO2Bulk}.
In the same plot, we show the initial fraction of Mu extracted
directly using $F^0_\mathrm{Mu}=2A_\mathrm{Mu}/A_\mathrm{tot}$~\cite{MuSiO2powder}.
%
As one can see, these values differ from the ones obtained
indirectly from $A_\mu$.
This is because the direct method is sensitive only to the fraction of Mu
that does not undergo fast relaxation, e.g., due to spin exchange
collisions in the pores \cite{MuDepolarizationSiO2}.
Note that because of the large gyromagnetic ratio Mu is much more
sensitive to depolarizing sources than $\mu^+$.

\begin{figure}[h]
\includegraphics[width=0.45\textwidth]{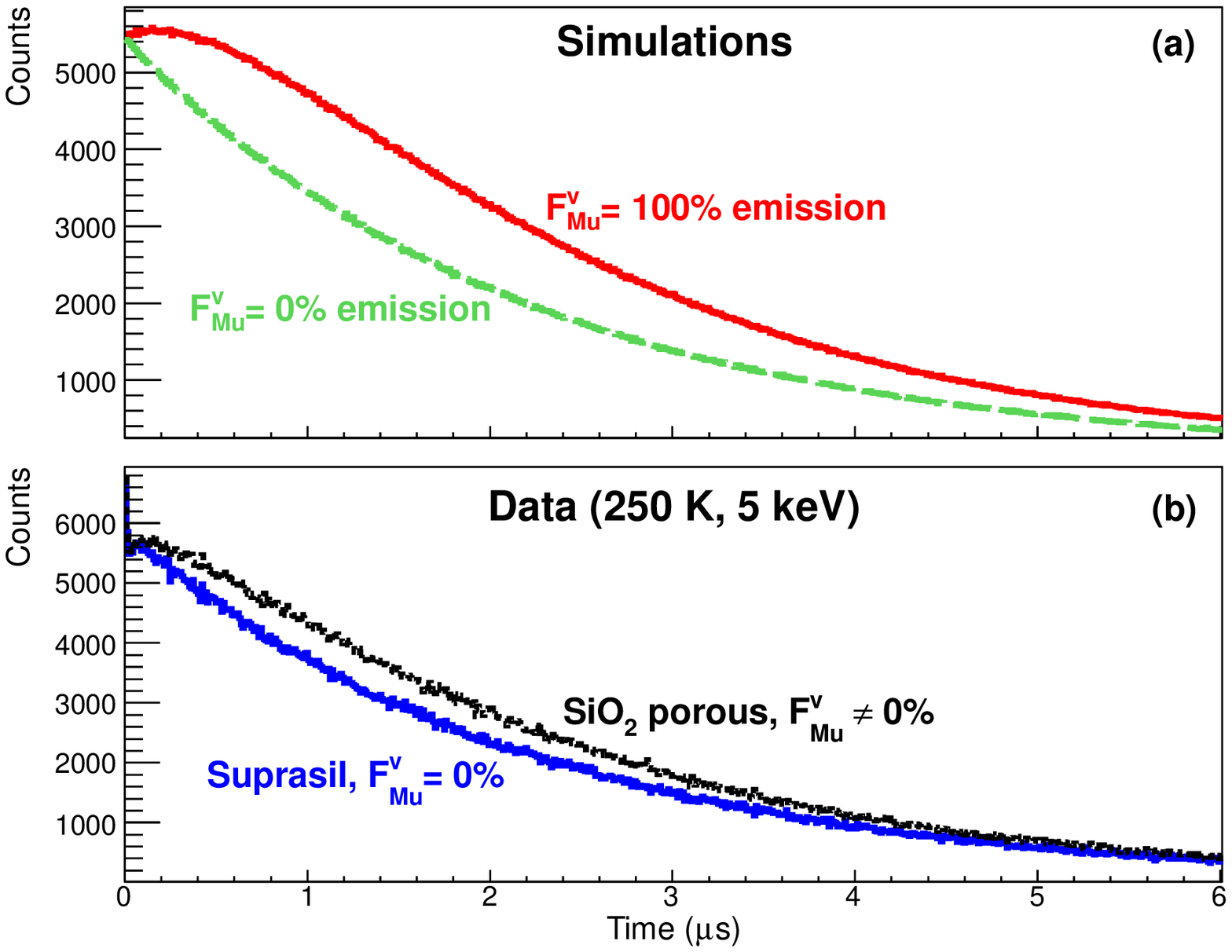}
\vspace{-0.7cm}
\includegraphics[width=0.43\textwidth]{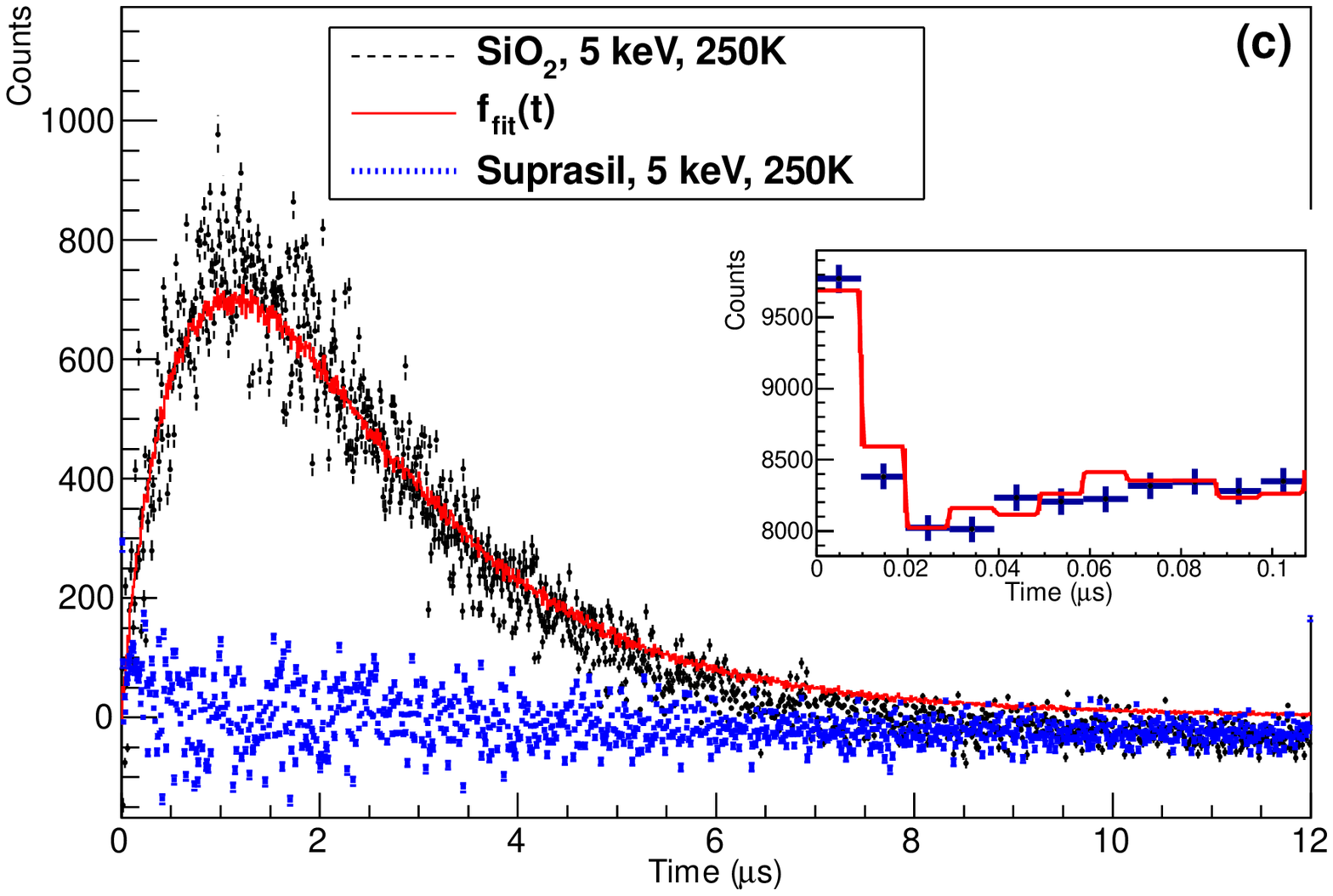}
\vspace{0.7cm}
\caption{(a) Simulated time distributions in the downstream detector
  for 0\% (dashed green) and 100\% (solid red).  (b) Measured time
  spectra for the porous material (dotted black) and the Suprasil sample
  (solid blue).  (c) Data and $f_\mathrm{fit}(t)$ after subtraction of the properly normalized exponential muon decay distribution. The
  inset shows the prompt peak and $f_\mathrm{fit}(t)$ (without subtraction).  }
\label{fig:timefit}
\end{figure}

Using the standard $\mu$SR setup allows us to determine the probability to form Mu. However with this technique we are unable to demonstrate Mu
emission into vacuum.
One possibility would be to use a tracking detector as
in~\cite{MuTracking}. We developed a new approach which exploits
the existing $\mu$SR setup.
The principle is based on the fact that the detection efficiency in
the downstream detectors (see Fig. \ref{setup}) is time dependent in
case of Mu emission into vacuum. Positrons from Mu decaying outside of
the film have a higher probability to be detected in the downstream
counters than the ones coming from $\mu^+$/Mu decays in the sample
which are shielded by the copper sample support.
Therefore, if vacuum emission occurs, a deviation from the $\mu^+$ exponential decay distribution is expected.
Hereafter, we will refer to this method as positron shielding
technique (PST).
Note that in PST we do not consider top, bottom, left and
right counters separately as in the $\mu$SR setup, but we only distinguish between upstream and downstream detectors.
In Fig.~\ref{fig:timefit} (a), we show the time spectra expected in
the downstream counters from simulations using
Geant4~\cite{geant4} for 0\% ($f_0$) and 100\% ($f_{100}$) Mu yield in
vacuum.
In Fig.~\ref{fig:timefit} (b), we present the measured data for
Suprasil (no emission into vacuum, thus corresponding to
0\%) and for SiO$_2$ porous material where emission into
vacuum is expected.
In order to determine the fraction of Mu emission into vacuum
($F\mathrm{^v_{Mu}}$), we fit the measured time spectra with
\begin{eqnarray*}
f_\mathrm{fit}(t)=n[(1-F\mathrm{^v_{Mu}})f_{0}(t) + F\mathrm{^v_{Mu}}f_{100}(t)]+
n_{pp}f_{pp}(t)
\end{eqnarray*}
where $n$ is the normalization and $n_{pp}f_{pp}(t)$ accounts
for a prompt peak.
This prompt peak which occurs in the first bins of the time spectra (see inset of
 Fig.~\ref{fig:timefit} (c)) originates from $\mu^+$ decaying in flight before
reaching the target and from back-scattered $\mu^+$\cite{MuonsTOF}. The
time distribution of this peak $f_{pp}(t)$ is determined experimentally using the Suprasil sample.
The three free parameters of the fit are $n$, $n_{pp}$ and $F\mathrm{^v_{Mu}}$.
Fits of $f_\mathrm{fit}(t)$ to the experimental data which have been taken for various implantation energies and film temperatures typically give a reduced 
$\chi^2$ of 1.1-1.4\, (612 degrees of freedom). 
In the simulations, Mu is assumed to be emitted from the surface of the
sample with a $\cos{\theta}$ angular
distribution~\cite{JanissenPhysRevA,cassidy} and an energy spectrum
corresponding to a Maxwell-Boltzmann distribution at the target
temperature.
Fitting the data with an isotropic angular distribution or a different temperature worsens the reduced $\chi^2$ by more than 0.2.

\begin{figure}[t!]
\includegraphics[width=0.4\textwidth]{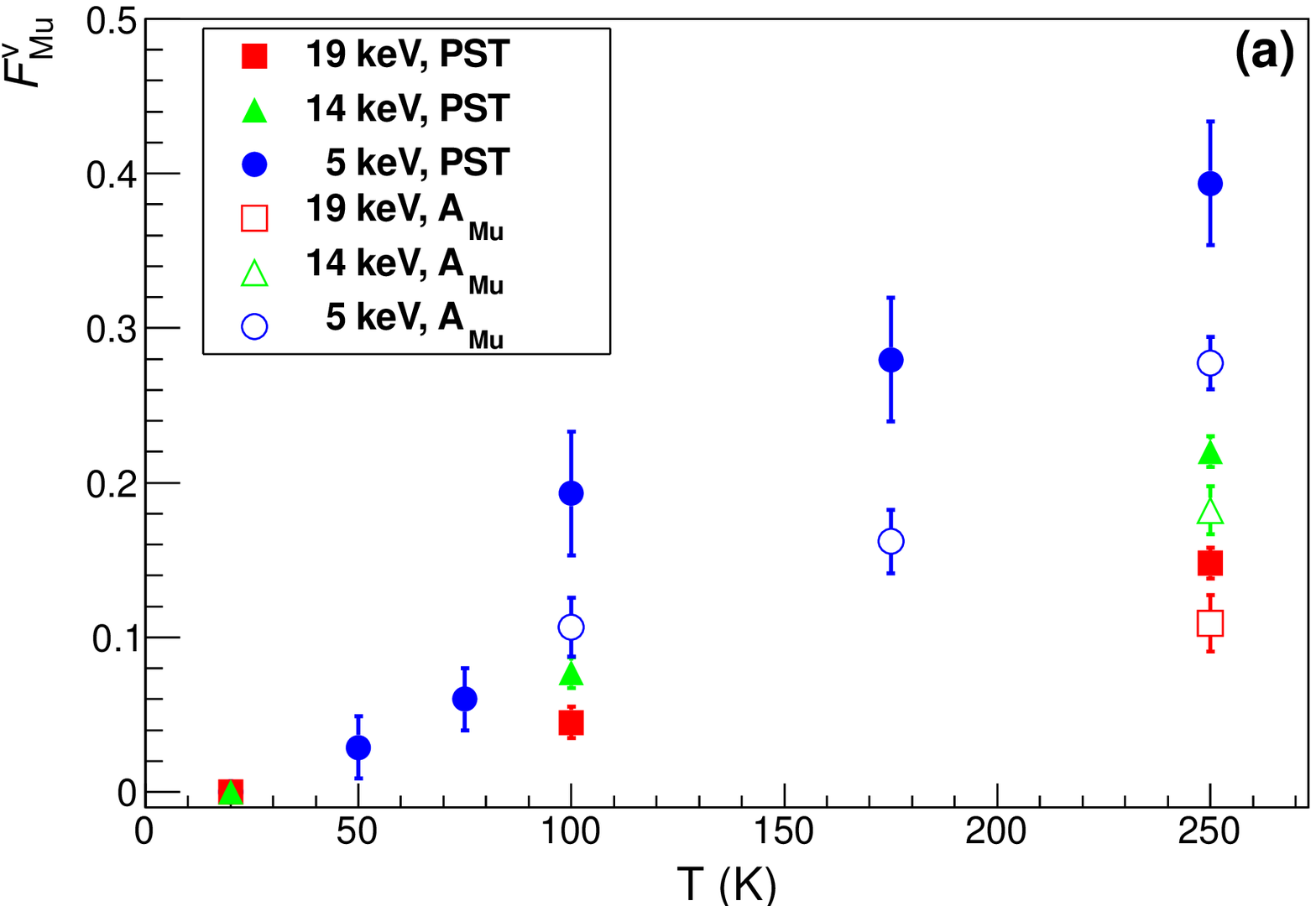}
\includegraphics[width=0.42\textwidth]{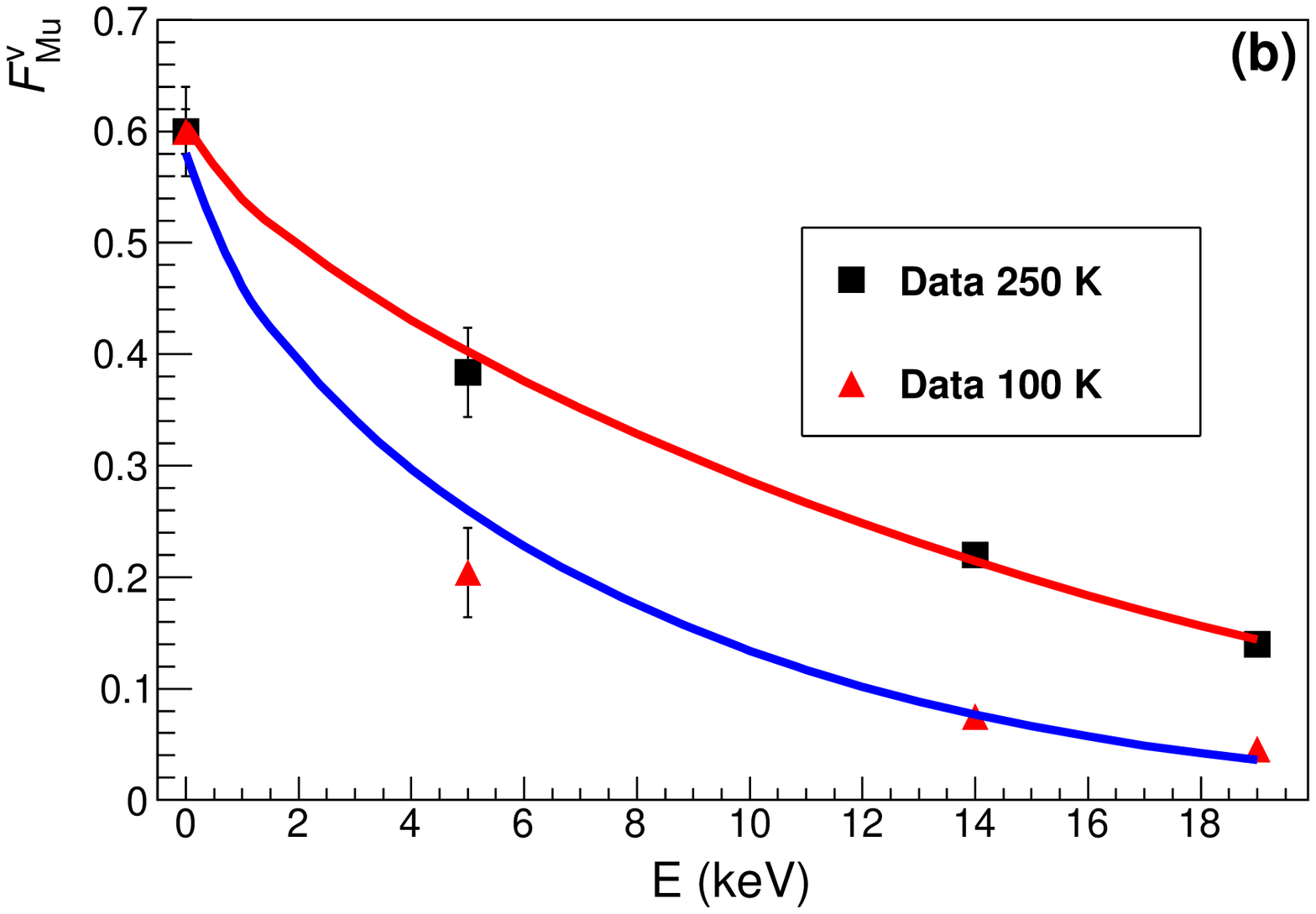}
\caption{(a) Vacuum yield $F\mathrm{^v_{Mu}}$ versus temperature determined with PST.
 For comparison, we show the results of the direct method. (b) $F\mathrm{^v_{Mu}}$ versus the implantation energy. The curves are fit with the diffusion model described in the text.}
\label{fig:YieldvsT}
\end{figure}

In order to better visualize the comparison between simulations and measurements,
in Fig.~\ref{fig:timefit} (c), we show the time spectrum after
subtraction of the prompt peak and the exponential muon decay distribution.
The Suprasil data give a constant value as
expected due to the absence of Mu emission from this sample.
%
%
On the contrary, for the porous film there is a clear signal caused by the
increased positron detection efficiency when Mu is emitted into vacuum.
The values of $F\mathrm{^v_{Mu}}$ extracted from the fits are
presented in Fig.~\ref{fig:YieldvsT}.
We obtain a yield of Mu into vacuum of $F\mathrm{^v_{Mu}}=(38\pm4)$\% at 250~K and
$F\mathrm{^v_{Mu}}=(20\pm4)$\% at 100~K for 5~keV implantation  energy.
The abrupt change of $F\mathrm{^v_{Mu}}$ visible between 75~K and 100~K
is due to thermal absorption of Mu at the pore walls as already
reported for silica powders~\cite{kieflHFS,harshawHFS}.
For 20 K $F\mathrm{^v_{Mu}}$ is compatible with zero.
The linear dependence of $F\mathrm{^v_{Mu}}\propto T$ 
(between 100~K and 250~K) is interesting since from a classical
diffusion model a $\sqrt{T}$-dependence is expected.
For comparison in Fig.~\ref{fig:YieldvsT} (a), we present also the
fraction of polarized Mu determined directly from the measurement of the Mu
asymmetry $A_\mathrm{Mu}$ with $\mu$SR technique.
As one can see, these points are systematically lower than $F\mathrm{^v_{Mu}}$ obtained with PST.
This is because PST, in contrast to the $\mu$SR direct method, is also sensitive to the fraction of Mu that depolarizes fast.
Nevertheless, both methods give consistent results in terms of dependence on the sample temperature and $\mu^+$ implantation energy ($E$).

In Fig.~\ref{fig:YieldvsT} (b),  $F\mathrm{^v_{Mu}}$ versus $E$ at 100 K and 250 K is fitted with a one dimensional diffusion model originally developed for Ps
\cite{PsDiffusion2,PsDiffusion4}.
The Mu fraction diffusing into vacuum is given by
$F\mathrm{^v_{Mu}}(E)=F^0_\mathrm{Mu}(E)J(E)$ with
$J(E)=\int_0^{l}e^{-\beta x}P(x,E)dx$,  where $l$ is the film
thickness, $\beta=1/\sqrt{D_\mathrm{Mu} \tau}$ the inverse of the diffusion
length and $D_\mathrm{Mu}$ the diffusion coefficient.
For the initial Mu fraction $F^0_\mathrm{Mu}$ we used the point at $E=0$ from Fig.~\ref{muSR}.
The $\mu^+$ implantation profile $P(x,E)$ was calculated using the
TrimSP simulation validated for $\mu^+$ with experimental
data~\cite{TrimSP}.
The only fit parameter to the data is the Mu diffusion constant
$D_\mathrm{Mu}$. 
The resulting values determined from the fits (solid
lines in Fig.~\ref{fig:YieldvsT} (b)) are $D\mathrm{^{250K}_{Mu}}=(1.6\pm0.1) \times
10^{-4}$~cm$^2$/s and $D\mathrm{^{100K}_{Mu}}=(4.2\pm0.5) \times
10^{-5}$~cm$^2$/s.
The good agreement between fit and data implies that $D_\mathrm{Mu}$ does not depend on the implantation
energy.
This means that the Mu thermalization time is much shorter than the
diffusion time (cfr. this result with similar measurements in
Ps, see Fig. 10 of~\cite{cassidy}).
A further argument that Mu quickly thermalizes is given by the
worsening of the $\chi^2$ when fitting the data of
Fig.~\ref{fig:timefit} (c) with distributions simulated at
temperatures different from the sample temperature.
Therefore, we can write the diffusion coefficient as a function of the mean
kinetic energy $E_\mathrm{Mu}$ of thermalized Mu in the pores
as $D_\mathrm{Mu}=\Lambda/(3 C) \sqrt{2 E_\mathrm{Mu}/m_\mathrm{Mu}}$
where $C$ is the mean number of collisions that Mu undergoes in one
pore before reaching the next one, $m_\mathrm{Mu}$ the Mu mass and
$\Lambda$ the mean distance between the pores~\cite{cassidy}.
Assuming an hexagonal close packing of the pores, one can estimate the
mean separation between them using  $\rho=\rho_0 (1-\pi
d^3/\Lambda^3 \sqrt{18})$~\cite{cassidy}.
We obtain $\Lambda=5.6$~nm for a pore diameter
of $d=5$~nm, a silica bulk density of $\rho_0=2.2$ g/cm$^3$ and a
porous film density of $\rho=1.1$ g/cm$^3$.
Using the experimentally determined $D_\mathrm{Mu}$, the mean number of
collisions in each pore is $C=2100\pm500$ at 100~K and
$C=850\pm100$ at 250~K.
These values confirm that Mu thermalization is fast ($\sim$ ns) on time scale of the diffusion process ($\sim$ $\mu$s). In fact, from the mass difference of Mu and SiO$_2$~\cite{ford1976}, one expects that in order to reach thermal energy Mu needs $\sim$500 collisions. %
%

The obtained $D_\mathrm{Mu}$ values are three orders of
magnitude smaller than expected from a classical diffusion
model~\cite{MuSiO2powder}.
In order to explain this disagreement, we interpret the Mu diffusion process  in the porous material
as quantum mechanical tunneling from pore to pore through a step potential barrier of (0.6$\pm$0.2)~nm width 
(corresponding to the pore walls thickness).
%
%
From the mean number of collisions $C$, which is the inverse of
the pore-to-pore tunneling probability, we deduce a height of the
potential barrier of (0.3$\pm$0.1)~eV. The uncertainty is dominated by our poor knowledge of the material structure. 
With this quantum mechanical model we can reproduce the observed
dependence of $D_\mathrm{Mu}(T)$ versus the temperature $T$. Since in our regime the
tunneling probability scales approximately linearly with $T$, we obtain that
$D_\mathrm{Mu}(T) \propto \sqrt{E_\mathrm{Mu}}/C \propto
\sqrt{T}/T^{-1}\propto T^{3/2}$.
The measured ratio $D\mathrm{^{250K}_{Mu}}/D\mathrm{^{100K}_{Mu}}\approx 3.8\pm0.5$ compares
well with the expected value from the $T^{3/2}$ dependence of (250K/100K)$^{3/2}\approx4$ supporting the validity of our model.
To check if the value of the potential barrier height obtained above can be identified with the Mu work-function ($W$), we performed Density Functional Theory calculations within GAUSSIAN 98~\cite{gaussian98} on clusters of SiO$_2$ containing up to eight silicon atoms and terminated by oxygen, capped with hydrogen atoms.
We compute the total energy $E^{\mathrm{tot}}_{\mathrm{SiO2+Mu}}$ of the SiO$_2$ matrix with a Mu atom
and the total energy of the SiO$_2$ fragment alone $E^{\mathrm{tot}}_{\mathrm{SiO2}}$. 
These computations of $W=E^{\mathrm{tot}}_{\mathrm{SiO2+Mu}}-E^{\mathrm{tot}}_{\mathrm{SiO2}}-13.6~\mathrm{eV}$ yield a value between -0.3 eV and -0.9 eV. The spread of the interval for $W$ originates from the uncertainty to locate the exact position of the interstitial Mu site with respect the Si and O atoms. Considering the over-simplification of our model, we conclude that our experimental determination of the work-function 
is consistent with the theoretical estimation. 
Further experiments using other
techniques and more precise measurements for Mu and Ps
will be useful to gain a deeper understanding of this intriguing
diffusion process in mesoporous films.

Summarizing, we have found that a sizeable fraction of thermalized Muonium is emitted into vacuum from mesoporous thin SiO$_2$ films.
At 250 K the yield is more than a factor two higher than previously found in SiO$_2$ powders at room temperature and comparable at 100 K. The high Muonium yield even at low temperatures is an important step towards the development of low emittance Mu sources for spectroscopy experiments.

This work was supported in part by the SNSF under the Ambizione grant PZ00P2\_132059, the SNFS grant
200021-138211 and the DOE Contract DE-FG02-07ER46352.  We thank A. Badertscher, U. Gendotti, F. Kottmann, R. Scheuermann,
D. Taqqu, the PSI and ETH workshops, the ETH Labortechnik group H. Scherrer, the PSI accelerator group and the NERSC and NU-ASCC computation centers.

\end{document}